\documentclass[12pt]{iopart}
\usepackage{graphicx}
\usepackage{iopams}
\usepackage[colorlinks=true,linktocpage=false,citecolor=blue]{hyperref}

\begin{document}
\title{Structure, Deformations and Gravitational Wave Emission of Magnetars}

\author{L Gualtieri, R Ciolfi and V Ferrari}
\address{Dipartimento di Fisica, ``Sapienza''
Universit\` a di Roma  and Sezione INFN Roma1,\\ 
piazzale Aldo Moro 2, I-00185 Roma, Italy}

%%%%%%%%%%%%%%%%%%%%%%%%%%%%%%%%%%%%%%%%%%%%%%%%%%%%%%%%%%%%%%%%%%%%%%%%%%%%%%%
%%%%%%%%%%%%%%%%%%%%%%%%%%%%%%%%%%%%%%%%%%%%%%%%%%%%%%%%%%%%%%%%%%%%%%%%%%%%%%%
\begin{abstract}
  Neutron stars can have, in some phases of their life, extremely strong magnetic fields, up to $10^{15-16}$ G. These
  objects, named magnetars, could be powerful sources of gravitational waves, since their magnetic field could determine
  large deformations. We discuss the structure of the magnetic field of magnetars, and the deformation induced by this
  field. Finally, we discuss the perspective of detection of the gravitational waves emitted by these stars.
\end{abstract}
\pacs{04.40.Dg, 04.30.Db}
%%%%%%%%%%%%%%%%%%%%%%%%%%%%%%%%%%%%%%%%%%%%%%%%%%%%%%%%%%%%%%%%%%%%%%%%%%%%%%% 
\section{Introduction}
%%%%%%%%%%%%%%%%%%%%%%%%%%%%%%%%%%%%%%%%%%%%%%%%%%%%%%%%%%%%%%%%%%%%%%%%%%%%%%% 
Magnetars are neutron stars (NSs) whose spin-down and bright emission activity are powered by the
stellar magnetic field. The interest of the scientific community in these objects has been growing since 1992, when
Tompson and Duncan \cite{DT92,TD93} first proposed a model which explains the spin-down rate and the emission properties
of two classes of astrophysical objects, the soft-gamma repeaters (SGRs) and the anomalous X-ray pulsars (AXPs) in terms
of strong magnetic fields. 

These objects have a very steep spin-down, and a very intense X-ray (and gamma-ray) activity, with periodic bursts of
$\sim10^{41}$ erg/s. Furthermore, in the last decades three giant flares from SGRs have been observed, with luminosities
reaching $\sim10^{47}$ erg/s. The observed spin-down of SGRs and AXPs corresponds (through the well-known dipole
emission formula $P\dot P\propto B^2$) to surface magnetic fields of the order of $10^{14}-10^{15}$ G. In the model of
Thompson and Duncan, the gamma activity is understood in terms of the evolution of the interior magnetic field, which is
as large as the surface field, or even larger; in their model, the field (or a significant fraction of it) is a toroidal
field\footnote{If we define a polar coordinate frame $(r,\theta,\phi)$ about the magnetic axis, the $r$- and
  $\theta$-components of the magnetic field are called {\it poloidal}, the $\phi$-component is called {\it
    toroidal}.}. This magnetic field has been produced in the early phases of the NS life, just after the supernova
explosion, due to flux conservation in the core collapse and/or to dynamo processes, related to convective motion and
differential rotation. It is worth noting that, although observed magnetars are slowly rotating, with periods $P$ of the
order of $10$ s, newly born magnetars could have much higher rotation rates, with periods $P\sim 10^{-3}$ s. Today we
know $18$ magnetars\footnote{For an up-to-date catalog, see {\tt
    http://www.physics.mcgill.ca/$\sim$pulsar/magnetar/main.html}}, but it is believed that a significant fraction
($\gtrsim10\%$) of NSs would possibly become magnetars at some stage of their evolution \cite{WT06}.

Due to their extreme properties, magnetars are very interesting objects both for astrophysics and for gravitational wave
physics. Quasi-periodic oscillations have been detected in the aftermath of the giant flares of SGRs; this is the first
observational evidence of NS oscillations \cite{qpo}. It has been suggested that magnetars may be the central engine for
some gamma-ray bursts \cite{DT92,grb,E02}. Last but not least, as we discuss below, the magnetic field could produce a
deformation much larger than that due to other mechanisms, thus magnetars are also interesting sources of gravitational
waves \cite{CJ}-\cite{MCSSV10}. We also remark that the present and future observational properties of magnetars could
shed light on the internal composition of NSs, and thus on the behaviour of matter at supranuclear densities.

For these reasons, in the last decades magnetars have been widely studied. However, their internal structure is still
poorly understood. We do not know, for instance, how strong is the interior magnetic field, and whether the toroidal
components prevail on the poloidal ones; we do not know whether the field is mainly dipolar or the higher order
multipoles dominate. This information would be very important, to understand the astrophysical processes involving
magnetars, and to assess the relevance of these stars as gravitational wave sources.

The magnetar model proposed in \cite{DT92,TD93} is dynamical, and the magnetic field evolves from its birth to its decay
\cite{GR92,HRV08}, through different processes (ambipolar diffusion, Hall drift, Ohmic decay). However, in some phases
of the early life of a neutron star it is legitimate to describe a magnetar as a stationary object, using the ideal
magnetohydrodynamics (MHD) approximation, as we shall briefly explain.

Let us consider what happens when a strongly magnetized neutron star is born.
\begin{itemize}
\item In the first seconds after the supernova explosion, the proto-neutron star is a very complicate and dynamical
  object, with turbulent and convective motion, differential rotation, and (eventually unstable) oscillations. In this
  period dynamo processes amplify the stellar magnetic field.
\item After few (or few tens of) seconds, convective instability is suppressed, and the matter composing the star can be
  described by a single, perfect fluid with infinite conductivity (ideal MHD approximation). As shown by numerical
  simulations in the ideal MHD approximation \cite{BS04,BS06}, the fluid is likely to settle down to a stationary
  configuration on a dynamical timescale of the order of Alfv\'en's time ($t_A\sim 0.01-10$ s).
\item After few minutes, matter becomes superfluid and the crust forms; thus the ideal MHD approximation no longer
  applies. The magnetic field evolves on timescales $\sim t_{decay}$ of the order of thousands of years or more.
\end{itemize} 
We remark that as the crust forms, the magnetic field is likely to freeze in the stationary configuration reached in the
previous stage. Therefore, this configuration could be an appropriate description of the stellar magnetic field for
timescales $t_A\lesssim t\lesssim t_{decay}$.

%%%%%%%%%%%%%%%%%%%%%%%%%%%%%%%%%%%%%%%%%%%%%%%%%%
\begin{figure}[!ht]
\begin{center}
\includegraphics[scale=.38]{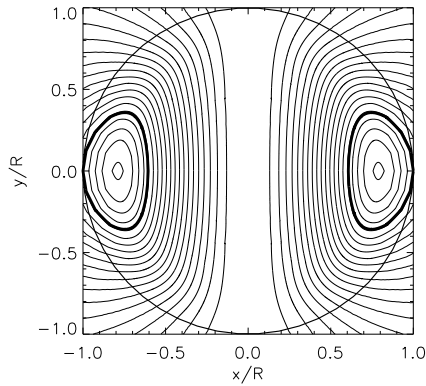}
\end{center}
\caption{The field lines of a twisted-torus magnetic field configuration, projected in the meridional plane. The
  toroidal field is confined in the region inside the thick curves.}
\label{tt}
\end{figure}
%%%%%%%%%%%%%%%%%%%%%%%%%%%%%%%%%%%%%%%%%%%%%%%%%% 
In the last decade many authors have been developing models of stationary magnetized neutron stars in ideal MHD
\cite{BG}-\cite{LJ09},\cite{CFGP08},\cite{CFG10}, including more and more ingredients in order to capture the essential
features of the system: poloidal and toroidal fields, general relativity, ``realistic'' equation of state (EOS). In
recent papers on the subject \cite{YYE06}-\cite{LJ09}, \cite{CFG10} a {\it twisted-torus configuration} has been
considered, in which the poloidal magnetic field extends throughout the star and in the exterior, whereas the toroidal
field is confined into a torus-shaped region inside the star, where the field lines are closed (see
Fig. \ref{tt}). There are different reasons for this choice:
\begin{itemize}
\item It has long been known that purely toroidal and purely poloidal magnetic field configurations are unstable
  \cite{T73,FR77,MRA10}; it is expected that a stable configuration should have both components \cite{P56}, with comparable amplitudes.
\item Numerical simulations \cite{BS04,BS06} have shown that the magnetic field tends to a twisted-torus configuration
  in which the toroidal and poloidal components have comparable amplitudes, for quite generic initial conditions (see
  also the analysis of \cite{B09}). This configuration appears to be stable, at least on a timescale $t_A\lesssim t\ll
  t_{decay}$. We remark that these simulations have been performed in a Newtonian framework, assuming a polytropic EOS
  for the stellar fluid.
\item The results of \cite{BS04,BS06} can be understood, at least qualitatively, as follows. Let us consider the {\it
    magnetic helicity}
\begin{equation}
H_m=\int{\bf A}\cdot{\bf B}\,dV
\end{equation}
where ${\bf A}$, ${\bf B}$ are the vector potential and the magnetic field, respectively (note that magnetic helicity
can also be defined in a relativistic framework). The following properties hold.
\begin{itemize}
\item The magnetic helicity is conserved on a timescale $\ll t_{decay}$. 
\item It vanishes if the field is either purely poloidal or purely toroidal. Thus, if the field is mixed (poloidal and
  toroidal) at the beginning, it must remain mixed for a long time.
\item The toroidal field is proportional to the electric current, thus, neglecting the stellar magnetosphere, it must
  vanish outside the star. \item The ratio between the toroidal and poloidal amplitudes can be described by a function
  $\zeta$, which is constant
  along each field line \cite{IS}. Therefore, a field line which extends outside the star must have $\zeta=0$, i.e. it
  must be purely poloidal.
\end{itemize}
It follows that, as the magnetic field reaches a stationary configuration, it must retain a mixed character, and 
the toroidal field must be confined inside the star, since the field lines with a non-vanishing toroidal component 
cannot cross the stellar surface. Such lines cover a torus-shaped region, tangent to the stellar surface at
the equator. This is the twisted-torus configuration (see Fig. \ref{tt}).
\end{itemize}
In the next Sections we discuss the features of magnetars with twisted-torus magnetic fields; our study is based on a
model we have recently developed \cite{CFGP09,CFG10} (see also \cite{CFGP08}). In Section \ref{configurations} we
briefly describe our model, and determine the magnetic field structure, discussing the relative amplitude of toroidal
and poloidal fields we expect. In Section \ref{deformations} we determine the stellar deformation induced by the
magnetic field, discussing how it depends on the EOS of the matter composing the star. In Section
\ref{gravitationalwaves} we discuss the possible gravitational emission of magnetars.
%%%%%%%%%%%%%%%%%%%%%%%%%%%%%%%%%%%%%%%%%%%%%%%%%%%%%%%%%%%%%%%%%%%%%%%%%%%%%%% 
\section{Structure}\label{configurations}
%%%%%%%%%%%%%%%%%%%%%%%%%%%%%%%%%%%%%%%%%%%%%%%%%%%%%%%%%%%%%%%%%%%%%%%%%%%%%%% 
We consider (see \cite{CFGP09,CFG10} for more details) a stationary, axisymmetric magnetized NS in the framework of general
relativity. We neglect stellar rotation (note that, as shown in \cite{YYE06}, twisted-torus configurations are not
significantly affected by stellar rotation) and the effect of the magnetosphere. Furthermore, we assume that the stellar
matter is described by a single perfect fluid with infinite conductivity (ideal MHD approximation). The magnetic field
is treated as a perturbation of a spherically symmetric background with metric
\begin{equation}
ds^{2}=-e^{\nu(r)}dt^2+e^{\lambda(r)}dr^2+r^2(d\theta^2+\sin^2\theta d\phi^2)
\end{equation}
($\nu,\lambda$ solutions of the unperturbed Einstein's equations describing the stellar structure) and four-velocity
$u^{\mu}=(e^{-\nu/2},0,0,0)$. We choose two EOSs, named APR2 \cite{APR} and GNH3 \cite{GNH3}, to model stars with large
and small compactnesses, respectively; indeed, a NS with mass $M=1.4\,M_\odot$ has radius $R=11.58$ km (APR2 EOS) or
$R=14.19$ km (GNH3 EOS).

The background is perturbed by a stationary, axisymmetric electromagnetic tensor $F_{\mu\nu}=A_{\mu,\nu}-A_{\nu,\mu}$,
associated to a current $j^\mu$, an electric field $E_\mu=F_{\mu\nu}u^\nu$ and a magnetic field
$B_\alpha=\frac{1}{2}\epsilon_{\alpha\beta\mu\nu}u^\beta F^{\mu\nu}$. The equations of ideal MHD are the baryon number
conservation $(nu^\mu)_{;\mu}=0$ ($n$ baryon density), the relativistic Euler equation $(\rho+p)a_\mu+p_{,\mu}+u_\mu
u^\nu p_{,\nu}-f_\mu=0$ ($\rho$ mass-energy density, $p$ pressure, $f_\mu=F_{\mu\nu}j^\nu$ Lorentz force), and the
vanishing of the electric field $E_\mu=0$.

With an appropriate gauge choice, the vector potential can be written as 
\begin{equation}
A_\mu=(0,e^{(\lambda-\nu)/2}\Sigma,0,\psi)
\end{equation}
where the ``flux function'' $\psi(r,\theta)$ describes the poloidal field, and the function $\Sigma(r,\theta)$ describes
the toroidal field. Neglecting higher order terms in the perturbation ($O(B^4)$), a remarkable property holds: the
quantity $\sin\theta\Sigma_{,\theta}$ only depends on the flux function $\psi$ (which is constant along each field
line). We can then define a function $\beta$ as
\begin{equation}
\sin\theta\Sigma_{,\theta}\equiv\beta=\beta(\psi)\,.
\end{equation}
Then, once we impose a form for $\beta(\psi)$, the magnetic field configuration is entirely determined by the flux
function $\psi(r,\theta)$, which can be found by solving the {\it relativistic Grad-Shafranov equation}:
\begin{eqnarray}
&&-\frac{e^{-\lambda}}{4\pi}\left[\psi''+\frac{\nu'-\lambda'}{2}\psi'
\right] -\frac{1}{4\pi r^2}\left[\psi_{,\theta\theta}-\cot{\theta}
\psi_{,\theta} \right] 
\nonumber\\
&&-\frac{e^{-\nu}}{4\pi}\beta\frac{d\beta}{d\psi}=(\rho+P)r^2\sin^2{\theta} [c_0+c_1\psi] 
\label{genGS}
\end{eqnarray}
with $c_0,c_1$ arbitrary constants. This equation follows from the ideal MHD equations.
By expanding the flux function $\psi(r,\theta)$ in Legendre polynomials as
\begin{equation}
\psi(r,\theta)=\sum_{l=1}^\infty a_l(r)\sin\theta P_{l,\theta}(\cos\theta)\,,
\end{equation}
Eq.~(\ref{genGS}) gives a coupled system of ordinary differential equations for the functions
$\{a_l(r)\}_{l=1,2,\dots}$. These equations admit two particular sets of solutions: the symmetric (with respect to the
equatorial plane) solutions, with vanishing even-order components ($a_{2l}\equiv0$) and the antisymmetric solutions,
with vanishing odd-order components ($a_{2l+1}\equiv0$). It is reasonable to expect that the actual field configuration
of these stars is, with a good approximation, symmetric with respect to the equatorial plane. Indeed, the magnetic field
has a nonvanishing dipole ($l=1$) component outside the star, and the antisymmetric solutions have vanishing magnetic
helicity, therefore symmetric solutions are energetically favoured with respect to the others. Furthermore, an
antisymmetric solution would likely be unstable on a dynamical timescale, since two opposite magnetic field loops could
annihilate each other. This would be in some sense similar to the Flowers-Ruderman instability of purely poloidal fields
\cite{FR77} (see also \cite{E02,MRA10}).

The twisted-torus configurations are those for which $\beta(\psi)$ is continuous and  has the form 
\begin{equation}
\beta(\psi)\sim\Theta(|\psi/\bar\psi|-1)\,,\label{betatt}
\end{equation}
where $\bar\psi\equiv\psi(R,\pi/2)$ is the value of the function $\psi$ on the stellar surface at the equator, and
$\Theta$ is the Heaviside step function. This can be understood by looking at Fig. \ref{tt}. The magnetic field lines
are also lines of constant $\psi$, and the thick line corresponds to $\psi=\bar\psi$. The toroidal region inside the
thick line has $\psi>\bar\psi$, and Eq.~(\ref{betatt}) implies that $\beta\neq0$, i.e. the toroidal field is
non-vanishing only in this region.

In \cite{CFG10} we solved the relativistic Grad-Shafranov equation, expanded in Legendre polynomials (with $l$ odd),
assuming a quite general parametrization for $\beta(\psi)$ (compatible with the twisted-torus condition (\ref{betatt}))
and employing the two EOSs APR2 and GNH3, which span a wide range of stellar compactness. A remarkable result we have
found is that the toroidal field never contributes to more than $13\%$ of the total magnetic energy of the star. This is
due to the fact that, if we enhance the amplitude of the toroidal field (roughly speaking, by making $\beta$ larger),
the region where the toroidal field is non-vanishing shrinks. Note however that, in this region, the toroidal field can
be larger than the poloidal field. Similar results have been obtained in \cite{LJ09}, using a polytropic EOS in a
Newtonian framework. We remark that this result, if confirmed, would challenge an assumption often used in magnetar
models \cite{TD96,SSIV05}, i.e. that the toroidal field prevails onto the poloidal inside the star.

The main open issue regarding these configurations is their stability. Indeed, they are stationary by construction, but
may be unstable. Actually, in \cite{B09} it has been found that magnetic field configurations in which the toroidal
field accounts for less than $20\%$ of the total magnetic energy appear to be unstable (in the framework of Newtonian
gravity and assuming a polytropic EOS). However, recent stability analyses of purely poloidal magnetic field
configurations (see \cite{LJ10} and references therein) show that the onset of the instability is localized along the
``neutral line'', which is the circle in the equatorial plane threading the closed field lines inside the star (see
Fig. \ref{tt}); as argued in \cite{LJ10}, a strong toroidal component along this line, like in the twisted-torus
configurations, could suppress the instability even when the overall energy of the toroidal field is small.
%%%%%%%%%%%%%%%%%%%%%%%%%%%%%%%%%%%%%%%%%%%%%%%%%%%%%%%%%%%%%%%%%%%%%%%%%%%%%%% 
\section{Deformations}\label{deformations}
%%%%%%%%%%%%%%%%%%%%%%%%%%%%%%%%%%%%%%%%%%%%%%%%%%%%%%%%%%%%%%%%%%%%%%%%%%%%%%% 
Once the magnetic field configuration has been determined with the perturbative approach outlined above, it is possible
to compute (perturbatively) the corresponding stellar deformation by solving Einstein's equations $\delta G_{\mu\nu}=
\frac{8\pi G}{c^4}\delta T_{\mu\nu}$ \cite{CFGP08,CFG10}.
The quadrupole ellipticity
\begin{equation}
\epsilon_Q=\frac{Q}{I}
\end{equation}
($Q$ mass-energy quadrupole moment, $I$ mean momentum of inertia) is the most relevant quantity encoding the stellar
deformation: it depends on the distribution of matter throughout the entire star (note that the gravitational wave
emission depends on $\epsilon_Q$). The mass-energy quadrupole moment $Q$ can be extracted by the far field limit of the
metric
\begin{equation}
g_{00}\rightarrow\dots-2Q\frac{e^\nu}{r^3}P_2(\cos\theta)
\end{equation}
and in the weak field limit it reduces to $Q\simeq\int_V\rho(r,\theta)r^2P_2(\cos\theta)dV$.

The poloidal field tends to make the star oblate, which corresponds to $\epsilon_Q>0$. The toroidal field, instead,
tends to make it prolate, i.e. with $\epsilon_Q<0$. The determination of the sign of $\epsilon_Q$ is important, because
if $\epsilon_Q<0$ a ``spin flip'' mechanism, suggested by Jones and Cutler \cite{CJ}, could take place: the angle
between the rotation axis and the magnetic axis would grow until they become orthogonal. This process would be
associated to a large gravitational emission. However, as discussed in Section \ref{configurations}, the stationary
twisted-torus configurations seem to be mainly poloidal, and indeed the corresponding deformations always have
$\epsilon_Q>0$. Therefore, the twisted-torus configurations seem not to be compatible with the Jones-Cutler mechanism.

The stellar deformation induced by twisted-torus magnetic field configurations depends on the EOS: less compact stars
have larger deformations. Furthermore, if one changes the magnetic field configuration (i.e. changes the choice of
$\beta(\psi)$ satisfying (\ref{betatt}), see Section \ref{configurations}), the stellar deformation changes less than
$10\%$. Note that, since the poloidal and toroidal fields have competing effects and the poloidal field prevails, it
follows that larger toroidal fields correspond to smaller deformations.

It is possible to summarize the deformations of these magnetized NSs as follows:
\begin{equation}
\epsilon_Q\simeq k\left[\frac{B_{pole}(G)}{10^{16}}\right]^2\times 10^{-4}\,.\label{deform}
\end{equation}
($B_{pole}$ is the amplitude of the dipolar surface magnetic field at the stellar pole). Here $k$ is a coefficient which
depends on the EOS: $k\sim4$ for the APR2 EOS, $k\sim9$ for the GNH3 EOS. We do not expect other EOSs to give results
very different from these (the results of \cite{LJ09} for a polytropic EOS are also similar).

The ellipticities (\ref{deform}) are larger than the bounds derived in \cite{UCB00,HJA06} (see also \cite{HK09}),
$|\epsilon|\lesssim 10^{-5}-10^{-6}$, by evaluating the maximal strain that the crust can sustain. However, these bounds
do not apply necessarily to our case. Indeed, here we consider a fluid star which is deformed by the magnetic field
before the crust forms. In this scenario, the equilibrium configuration of the crust would be its initial, non-spherical
shape, and the limits derived in \cite{UCB00,HJA06} may be violated. However, we do not know how long the crust would
remain in a non-spherical shape: in order to understand the evolution and the persistence of the stellar deformation, a
dynamical study of the magnetic field evolution on longer time-scales would be needed.
%%%%%%%%%%%%%%%%%%%%%%%%%%%%%%%%%%%%%%%%%%%%%%%%%%%%%%%%%%%%%%%%%%%%%%%%%%%%%%% 
\section{Gravitational wave emission}\label{gravitationalwaves}
%%%%%%%%%%%%%%%%%%%%%%%%%%%%%%%%%%%%%%%%%%%%%%%%%%%%%%%%%%%%%%%%%%%%%%%%%%%%%%% 
If an axisymmetric NS with quadrupole ellipticity $\epsilon_Q$ induced by a magnetic field, rotates about an axis forming
an angle $\alpha$  with the magnetic axis, it emits gravitational waves. If $\alpha$ is small, 
gravitational radiation is mainly emitted at the same frequency $\nu$ as the rotation rate, with amplitude
\begin{equation}
h_0\simeq\frac{4G}{rc^4}(2\pi\nu)^2I|\epsilon_Q|\sin\alpha\,.\label{h0}
\end{equation}
We remark that the best available estimate of the ``wobble angle'' $\alpha$ of a neutron star is $\alpha=3^o$ for PSR
B1828-11 \cite{wobble}.  In the Jones-Cutler process, which takes place as $\epsilon_Q<0$, the wobble angle increases
towards $\alpha=90^o$, with a great enhancement of the gravitational radiation. However, as discussed in Section
\ref{deformations}, this is not the case for the twisted-torus configurations.

The detectability of gravitational emission from magnetically deformed NS, described by Eq.~(\ref{h0}), depends both
on the overall magnetic field amplitude, which determines $\epsilon_Q$, and on the duration of the emission
process. Indeed, different dissipative processes tend to reduce both the wobble angle and the rotation frequency, 
then reducing the time the emission frequency spends in the bandwidth of ground-based interferometers (from few
tens to few hundreds of Hertz).
\begin{itemize}
\item As discussed in \cite{JA}, the wobble angle of an oblate ($\epsilon_Q>0$) star with rotation period $P$ would decay, due to internal dissipation, in a
timescale
\begin{equation}
\tau_d\sim \frac{nP}{\epsilon_Q}
%32\,{\rm yr}\,\left(\frac{n}{10^4}\right) \left(\frac{100\,Hz}{\nu}\right) \left(\frac{I}{10^{45}\hbox{g\,cm}^2}\right)
\label{td}
\end{equation}
where the parameter $n$ is unknown, since we do not have a clear understanding of the damping processes. An estimate of
this parameter for slowly rotating stars has been proposed by Alpar \& Sauls \cite{AS}: $400\lesssim n\lesssim10^4$. This
would correspond, for instance, to a damping timescale ranging from few months to few years if $B_{pole}\sim 10^{15}$ G.
Therefore, after at most few years the rotation and symmetry axis would become nearly parallel, and the emission would
become negligible, unless some pumping mechanism \cite{JA} takes place which increases the wobble angle.
\item A NS with dipolar field at the pole $B_{pole}$ and wobble angle $\alpha$ spins down with a period derivative $\dot P$ 
given by \cite{WT06,pal}
\begin{equation}  
B_{pole}\simeq 6.4\times 10^{19}\frac{\sqrt{P\dot P}}{\sin\alpha}\,{\rm G}\label{spdwn}
\end{equation} 
(note that many authors consider the average surface magnetic field, which is $\frac{1}{2}B_{pole}$ \cite{NK10,ST}). Since
Eq.~(\ref{spdwn}) implies that $\dot P=K/P$ (where $K$ is a constant), the star slows down from an initial period
$P_{in}$ to a period $P_{fin}$ in the characteristic time
\begin{equation} 
\tau_c=\frac{1}{2K}(P_{fin}^2-P_{in}^2)\simeq\frac{P^2_{fin}}{2K}\,.\label{tc}
\end{equation} 
Therefore, a NS with $B_{pole}$ of the order of $10^{15}$ G and a small wobble angle could lie in the bandwidth of ground
based interferometers for a time ranging from few months to few years. If the magnetic field is larger the star spins
down more rapidly, making detection more difficult. The detection is also unlikely if the parameter $n$ in (\ref{td}) is much
smaller than the upper limit $\sim 10^4$, since in this case the wobble angle would rapidly decay.
\end{itemize}
%%%%%%%%%%%%%%%%%%%%%%%%%%%%%%%%%%%%%%%%%%%%%%%%%%
\begin{figure}[!ht] 
\begin{center} 
\includegraphics[angle=270,scale=.28]{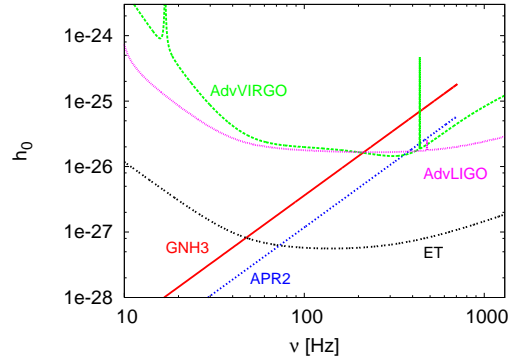} 
\end{center} 
\caption{Gravitational
    signal emitted by a rotating NS at a distance of $10$ kpc, deformed by a twisted-torus magnetic field, with
    $B_{pole}=10^{15}$ G, $\alpha=3^o$, $M=1.4\,M_\odot$ and EOS GNH3 (solid line) and APR2 (dotted line). The sensitivity
    curves of advanced LIGO, advanced VIRGO (with integration time of three months) and ET (with integration time of one
    year) are shown for comparison.}
\label{h0fig}
\end{figure}
%%%%%%%%%%%%%%%%%%%%%%%%%%%%%%%%%%%%%%%%%%%%%%%%%%
In Fig. \ref{h0fig} we show the signal emitted from a NS with $B_{pole}=10^{15}$ G and wobble angle $\alpha=3^o$, at a
distance of $10$ kpc (i.e. in our galaxy), computed by Eqns.~(\ref{deform}), ({\ref{h0}).  The star initially rotates
with $\nu=700$ Hz, which is close to the largest rotation frequency of known pulsars; note that magnetars are believed
to rotate rapidly at birth \cite{DT92}.  This signal is compared with the sensitivity curves of the advanced detectors
LIGO, VIRGO (assuming an integration time of three months) and of the third generation detector ET (assuming one year 
integration time)\footnote{{\tt http://www.ligo.caltech.edu; http://www.ego-gw.it; http://www.et-gw.eu}}. An estimate
of the spin-down time $\tau_c$ by Eq.~(\ref{tc}) shows that, if the wobble angle decay is not too fast, the signal
lies in the bandwidth of advanced LIGO/VIRGO for a few months, and it lies in the bandwidth of ET for a few years,
consistently with the integration times we have employed (see also \cite{F10}).

Figure \ref{h0fig} shows that the signal could be well detected by ET, and marginally detected by advanced LIGO/VIRGO.
However, one should also take into account the event rate of the process generating the gravitational wave signal. In
our scenario, a NS could maintain a strong, twisted-torus magnetic field and the corresponding deformation for several
years (in the most optimistic case, up to thousands of years), thus there may be several NS in our galaxy with large
deformation. However, only few of them would rotate rapidly enough to be detected by ground based interferometers. If
there is no spin-up process, the rate of the events described in Fig. \ref{h0fig} would be at most the same as NS birth
rate, i.e. few per century in our galaxy. On the other hand, accretion from a companion star could spin-up the star and
increase the wobble angle; in this case the event rate may be significantly larger. We also mention that, as discussed
in \cite{MCSSV10}, the stochastic background of gravitational waves from magnetars could be detectable by the third
generation detector ET.

Finally, we mention that LIGO and Virgo set an upper limit of the order of $\sim 10^{-4}$ on the deformations of known
pulsars (Crab, J0537-6910 and J1952+3252) \cite{LIGO}. Indeed, larger deformations would have produced signals strong
enough to be detected. These limits are stronger than current limits arising from spin-down \cite{pal}. We remark that
the deformations considered in the analysis of \cite{LIGO} are different from those considered here and in current
literature on magnetars. Indeed, we consider axially symmetric stars inclined by an angle $\alpha$ with respect to the
rotation axis (which yield gravitational waves at frequency $\nu$). In the data analysis carried on in \cite{LIGO},
instead, tri-axial deformations (without inclination) have been considered, which yield gravitational waves at frequency
$2\nu$.
%%%%%%%%%%%%%%%%%%%%%%%%%%%%%%%%%%%%%%%%%%%%%%%%%%%%%%%%%%%%%%%%%%%%%%%%%%%%%%%
\section*{Acknowledgements}
L.G. thanks Tsvi Piran for useful comments.  This work was partially supported by CompStar, a Research Networking
Program of the European Science Foundation. LG has been partially supported by the grant PTDC/FIS/098025/2008.
%%%%%%%%%%%%%%%%%%%%%%%%%%%%%%%%%%%%%%%%%%%%%%%%%%%%%%%%%%%%%%%%%%%%%%%%%%%%%%%

%%%%%%%%%%%%%%%%%%%%%%%%%%%%%%%%%%%%%%%%%%%%%%%%%%%%%%%%%%%%%%%%%%%%%%%%%%%%%%%
\end{document}